\documentclass[aps,twocolumn,showpacs,amsmath,amssymb]{revtex4}
\usepackage{graphicx}
\usepackage{bm}

\def\br{ \bm{r} }

\def\bk{ \bm{k} }
\def\bq{ \bm{q} }

\begin{document}

\title{CePt$_3$Si: an unconventional superconductor without inversion center}

\author{K. V. Samokhin, E. S. Zijlstra, S. K. Bose}

\address{Department of Physics, Brock University, St.Catharines, Ontario, Canada L2S 3A1}
\date{\today}

\begin{abstract}
Most superconducting materials have an inversion center in their
crystal lattices. One of few exceptions is the recently discovered
heavy-fermion superconductor CePt$_3$Si [E. Bauer {\em et al},
Phys. Rev. Lett. \textbf{92}, 027003 (2004)]. In this paper, we
analyze the implications of the lack of inversion symmetry for the
superconducting pairing. We show that the order parameter is an
odd function of momentum, and that there always are lines of zeros
in the excitation energy gap for one-component order parameters,
which seems to agree with the experimental data. The
superconducting phase can be non-uniform, even without external
magnetic field, due to the presence of unusual gradient terms in
the Ginzburg-Landau free energy. Also, we performed {\em ab
initio} electronic structure calculations for CePt$_3$Si, which
showed that the spin-orbit coupling in this material is strong,
and the degeneracy of the bands is lifted everywhere except along
some high symmetry lines in the Brillouin zone.
\end{abstract}


\maketitle

\section{Introduction}
\label{sec: Intro}

In the past two decades, a number of novel superconducting
materials have been discovered where order parameter symmetries
are different from an $s$-wave spin singlet, predicted by the
Bardeen-Cooper-Schrieffer (BCS) theory of electron-phonon mediated
pairing. From the initial discoveries of unconventional
superconductivity in heavy-fermion compounds, the list of examples
has now grown to include the high-$T_c$ cuprate superconductors,
ruthenates, ferromagnetic superconductors, and possibly organic
materials. In most of these materials, there are strong
indications that the pairing is caused by the electron
correlations, in contrast to conventional superconductors such as
Pb, Nb, {\em etc}. Non-phononic mechanisms of pairing are believed
to favor a non-trivial spin structure and orbital symmetry of the
Cooper pairs. For example, the order parameter in the high-$T_c$
superconductors, where the pairing is thought to be caused by the
anti-ferromagnetic correlations, has the $d$-wave symmetry with
lines of zeroes at the Fermi surface.

A powerful tool of studying unconventional superconducting states
is symmetry analysis, which works even if the pairing mechanism is
not known. Mathematically, superconductivity is a consequence of
the breaking of the gauge symmetry $U(1)$ in the full symmetry
group of the normal state: ${\cal G}=S_M\times U(1)$, where $S_M$
is a magnetic space group, which includes usual space group
operations (i.e. translations, rotations, inversion, {\em etc})
and time reversal operation $K$. In a non-magnetic crystal,
$S_M=S\times{\cal K}$, where $S$ is the space group, and ${\cal
K}=\{E,K\}$. In a magnetic crystal, $K$ enters $S_M$ only in
combination with other symmetry elements. The superconducting
state is said to be ``unconventional'' if, in addition to the
gauge symmetry, some other symmetries from ${\cal G}$ are broken.
The group-theoretical analysis of unconventional superconducting
states in non-magnetic crystals was developed in Refs.
\cite{VG85,UR85}. Recently, it was extended to include the
ferromagnetic case \cite{FMSC,FominFMSC,MineevFMSC}.

In almost all previous studies, it was assumed that the crystal
has an inversion center, which leads to degenerate electron bands
and makes it possible to classify the Cooper pair states according
to their spin (or pseudospin if the spin-orbit coupling is taken
into account). The even (singlet) and odd (triplet) components of
the superconducting order parameter can then be studied separately
\cite{Book}. Although this is the case in most superconductors,
there are some exceptions. Early discussion of the possible loss
of inversion symmetry associated with a structural phase
transition in V$_3$Si, which is an A15 type superconductor, can be
found in Ref. \cite{AB65}. Later, a C15 superconductor HfV$_2$ was
found to undergo a transition from a cubic to a
non-centrosymmetric body-centered orthorhombic structure
\cite{LZ72}. The possible existence of superconductivity was
reported in ferroelectric perovskite compounds SrTiO$_3$
\cite{SHC64} and BaPbO$_3$-BaBiO$_3$ \cite{BV83}. More recently,
there has been a renewed theoretical interest in this problem,
mainly in the context of superconductivity in two-dimensional (2D)
systems, such as Cu-O layers in YBCO \cite{Edelstein}, or surface
superconductivity on Tamm levels \cite{GR01,BG02}. According to
Ref. \cite{Edelstein,GR01}, in the absence of inversion symmetry
the order parameter becomes a mixture of spin-singlet and
spin-triplet components, which leads, for instance, to the Knight
shift attaining a non-zero value at $T=0$. Other peculiar
properties of non-centrosymmetric superconductors include a jump
of magnetic induction at the surface \cite{LNE85}, and the
possibility of non-uniform helical superconducting phases due to
the presence of first-order gradient terms in the Ginzburg-Landau
(GL) functional \cite{MS94}.

This work is motivated by a recent experimental discovery of
superconductivity with $T_c\simeq 0.75K$ in CePt$_3$Si
\cite{exp-CePtSi}, which is the first known heavy-fermion material
without inversion center. (It should be mentioned that
incommensurate density modulations might break inversion symmetry
in a heavy-fermion compound UPt$_3$, as pointed out in Ref.
\cite{Min93}.) Our goal is two-fold. First, we study the symmetry
of electron bands, calculate the electronic structure, and
estimate the magnitude of the spin-orbit coupling in the normal
state of CePt$_3$Si. Second, we use general symmetry arguments to
analyze the gap symmetry and spatial dependence of the order
parameter in a three-dimensional non-centrosymmetric tetragonal
superconductor, assuming a strong spin-orbit coupling and the
clean limit. The paper is organized as follows. In Sec. \ref{sec:
Symmetry of bands}, we study symmetry of the single electron
states. In Sec. \ref{sec: Bands}, the results of the electronic
band structure calculations are reported. Sec. \ref{sec: OP}
focuses on the symmetry of the superconducting order parameter,
possible locations of the gap nodes, and also spatial structure of
the superconducting phase, using the phenomenological
Ginzburg-Landau approach. Sec. \ref{sec: Concl} concludes with a
discussion of our results.

\section{Symmetry of electron bands}
\label{sec: Symmetry of bands}

The symmetry group of the normal paramagnetic state: ${\cal
G}=S\times{\cal K}\times U(1)$, where $S$ is the space group of
the crystal, and $U(1)$ is the gauge group. In the case of
CePt$_3$Si, the space group is P4mm (No. 99), which is generated
by (i) lattice translations by the primitive vectors
$\bm{a}=a(1,0,0)$, $\bm{b}=a(0,1,0)$, $\bm{c}=c(0,0,1)$ of a
tetragonal lattice, and (ii) the generators of the point group
$G=\mathbf{C}_{4v}$: the rotations $C_{4z}$ about the $z$ axis by
an angle $\pi/2$ and the reflections $\sigma_x$ in the vertical
plane $(100)$. Spatial inversion $I$ is not an element of the
symmetry group.

In the absence of inversion symmetry, spin-orbit (SO) coupling
plays a crucial role due to its band-splitting effect. At non-zero
SO coupling, the single-particle wavefunctions are linear
combinations of the eigenstates of the spin operator $s_z$:
$\langle\br|\psi\rangle=u(\br)\chi_\uparrow+
v(\br)\chi_\downarrow$. Since the normal state Hamiltonian $H_0$
is invariant with respect to the crystal lattice translations, the
eigenfunctions are the Bloch spinors $\psi_{\bk,n}(\br)$ belonging
to the wave vectors $\bk$ in the Brillouin zone, which can be
written in the form
\begin{equation}
\label{psi_pm}
  \langle\br|\bk,n\rangle=u_{\bk,n}(\br)\chi_\uparrow+
    v_{\bk,n}(\br)\chi_\downarrow.
\end{equation}
The corresponding eigenvalues $\epsilon_{n}(\bk)$ describe the
band dispersion of free electrons.

In the presence of both the time-reversal and inversion symmetries
the bands are two-fold degenerate at each $\bk$. Indeed, the
states $|\bk,n\rangle$ and $KI|\bk,n\rangle$ correspond to the
same $\bk$, have the same energy, and are orthogonal (in addition,
these two states are degenerate with another pair of orthogonal
states, $K|\bk,n\rangle$ and $I|\bk,n\rangle$, which correspond to
$-\bk$). In this case, $n=(\nu,\pm)$, where $\pm$ labels the
linearly independent Bloch states at a given band index $\nu$.
There is a freedom in choosing the basis functions
$|\bk,\nu,+\rangle$ and $|\bk,\nu,-\rangle$. The most frequently
used convention is that they should transform under rotations $R$
similar to the spin eigenstates $|\bk,\nu,\uparrow\rangle$ and
$|\bk,\nu,\downarrow\rangle$, i.e.
\begin{equation}
\label{pseudospin def}
     R|\bk,\nu,\alpha\rangle=
     U_{\alpha\beta}(R)|R\bk,\nu,\beta\rangle,
\end{equation}
where $\alpha,\beta=\pm$ and $U(R)$ is the spin-rotation matrix:
for a rotation by an angle $\theta$ around some axis $\bm{n}$:
$U(R)=\exp[-i(\theta/2)(\bm{\sigma}\cdot\bm{n})]$ ($\bm{\sigma}$
are Pauli matrices). The practical recipe for constructing the
basis of the Bloch states is as follows: first choose a state
$|\bk,\nu,+\rangle$ at each $\bk$ in the irreducible part of the
first Brillouin zone, then act on it by $KI$ to obtain an
orthogonal state $|\bk,\nu,-\rangle$ at the same $\bk$, and
finally act on $|\bk,\nu,\pm\rangle$ by the elements of the point
group and use the prescription (\ref{pseudospin def}) to obtain
the pairs of the basis functions belonging to the star of $\bk$.
The single-electron states constructed in this way are referred to
as the pseudospin states \cite{UR85}. It is the presence of the
inversion center that makes the bands double degenerate in a
non-magnetic crystal:
$\epsilon_{\nu,+}(\bk)=\epsilon_{\nu,+}(-\bk)=\epsilon_{\nu,-}(\bk)$
at all $\bk$.

In contrast, if the crystal lacks the inversion symmetry, then the
degeneracy of the single-electron bands $\epsilon_n(\bk)$ is
lifted everywhere, except from some points or lines of high
symmetry. In particular, the bands always touch at
$\bk=\mathbf{0}$ (the $\Gamma$ point), because of the Kramers
theorem: time-reversal symmetry means that $|\bk,n\rangle$ and
$K|\bk,n\rangle\sim|-\bk,n\rangle$ have the same energy. In the
case of CePt$_3$Si, one can show that the spin-split bands touch
along the $\Gamma Z$ line (i.e. along the [001] direction), and
there are no other symmetry-imposed degeneracies in the bands
crossing the Fermi level (see Sec. \ref{sec: Bands} below). In the
limit of zero SO coupling (which is not applicable to CePt$_3$Si),
but still without an inversion center, the symmetry of the system
contains $SU(2)$ spin rotations, in addition to the space group,
time reversal, and the gauge group. Then, the bands at each $\bk$
are double degenerate.

Let us now show that the transformation properties of the
single-electron wave functions in the absence of inversion center
are different from those of the spin or pseudospin eigenstates,
see Eq. (\ref{pseudospin def}). Mathematically, the wave functions
in each band transform according to irreducible co-representations
of the Type II magnetic space group $S_M=S\times{\cal K}$
\cite{co-reps} (one should use co-representations because the
time-reversal operation $K$ is anti-unitary). In addition, the
co-representations must be double-valued because the states
$|\bk\rangle$ are spin-$1/2$ spinors, so that any rotation by
$2\pi$ changes their sign: $C_{4z}^4|\bk\rangle=-|\bk\rangle$,
$\sigma_x^2|\bk\rangle=-|\bk\rangle$, and also
$K^2|\bk\rangle=-|\bk\rangle$ (the band index $n$ is omitted). The
double-valuedness can be dealt with in the standard fashion by
using the ``double-group'' trick \cite{LL-3}: one introduces a
fictitious new symmetry element $\bar E$, which commutes with all
other elements and satisfies the conditions
$C_{4z}^4=\sigma_x^2=\bar E$, $\bar E^2=E$, and also $K^2=\bar E$.

The co-representations of $S_M$ can be derived from the usual
representations of the unitary component, which in our case
coincides with the space group $S$. For each $\bk$ in the
Brillouin zone, the basis of an irreducible representation of $S$
is formed by the Bloch states corresponding to the star of $\bk$.
The state $K|\bk\rangle$ belongs to the wave vector $-\bk$, but
the irreducible representations of $S$ corresponding to $\bk$ and
$-\bk$ are inequivalent (because of the absence of inversion
symmetry) and must therefore be regarded as belonging to a single
``physically irreducible'' representation of twice the dimension
\cite{LL-3}. In terms of co-representations this means that the
Bloch states $|\bk\rangle$ and $K|\bk\rangle$ belong to the same
two-dimensional irreducible co-representation of $S\times{\cal K}$
\cite{co-reps}. Thus, the appropriate basis of the Bloch states is
formed by the wave functions $|G\bk\rangle$ corresponding to the
star of $\bk$, and also by their time-reversed counterparts
$K|G\bk\rangle$, which can be combined in a set of bispinors
$|\Psi_{\bk}\rangle=(|\bk\rangle,K|\bk\rangle)^T$. All the states
from this set have the same energy:
$\epsilon(\bk)=\epsilon(G\bk)=\epsilon(-\bk)$.

Since the function $G|\bk\rangle$ belongs to the wave vector
$G\bk$, one can write
$G|\bk\rangle=e^{i\varphi_{\bk}(G)}|G\bk\rangle$. The undetermined
phase factors come from the freedom in choosing the phases of the
Bloch states and realize a representation of the point group in
this basis (it is assumed that the Bloch states are single-valued
and continuous functions throughout the Brillouin zone). Using the
commutation of $G$ and $K$ we have $GK|\bk\rangle=KG|\bk\rangle=
K\exp[i\varphi_{\bk}(G)]|G\bk\rangle=
\exp[-i\varphi_{\bk}(G)]K|G\bk\rangle$. The co-representation
matrices in the basis of bispinor wave functions can be obtained
from the relations
\begin{eqnarray}
    \label{trans G}
    && G|\Psi_{\bk}\rangle=\left(\begin{array}{cc}
    e^{i\varphi_{\bk}(G)} & 0 \\
    0 & e^{-i\varphi_{\bk}(G)} \\
    \end{array}\right)|\Psi_{G\bk}\rangle\\
    \label{trans bE}
    && \bar E|\Psi_{\bk}\rangle=\left(\begin{array}{cc}
    -1 & 0 \\
    0 & -1 \\
    \end{array}\right)|\Psi_{\bk}\rangle\\
    \label{trans K}
    && K|\Psi_{\bk}\rangle=
    \left(\begin{array}{cc}
    0 & 1 \\
    -1 & 0 \\
    \end{array}\right)|\Psi_{\bk}\rangle
\end{eqnarray}
[we have taken into account that $K(K|\bk\rangle)=-|\bk\rangle$].
Note that the multiplication rules for the co-representation
matrices are different from usual unitary group representations.
For example, $D(G_1G_2)=D(G_1)D(G_2)$ and $D(GK)=D(G)D(K)$, but
$D(KG)=D(K)D^*(G)$ and $D(K^2)=D(K)D^*(K)$ \cite{co-reps}.

The general symmetry arguments given above can be illustrated
using an exactly-solvable three-dimensional generalization of the
Rashba model, which was originally proposed to describe the
effects of symmetry lowering near the surface of a semiconductor
\cite{Rashba60} and recently applied in Refs.
\cite{Edelstein,GR01} to quasi-2D non-centrosymmetric
superconductors. Consider a single band $\epsilon_0(\bk)$ in a
crystal described by the point group $\mathbf{C}_{4v}$. At zero SO
coupling, the band is two-fold degenerate due to spin. The absence
of reflection symmetry in the $xy$ plane implies the existence of
an internal electric field in the crystal, whose average over a
unit cell is non-zero. In the Rashba approximation, this
non-uniform field is replaced by its average, introducing a
constant vector $\bm{n}\parallel\hat z$. When a non-zero SO
coupling is switched on, it can described by an additional term in
the Hamiltonian:
\begin{equation}
\label{Rashba}
    H_{so}=\alpha\sum\limits_{\bk}\bm{n}\cdot(\bm{\sigma}_{\sigma\sigma'}
    \times\bk)\;a^\dagger_{\bk,\sigma}a_{\bk,\sigma'},
\end{equation}
where $\sigma,\sigma'=\uparrow,\downarrow$ is the $z$-axis spin
projection, and the states $|\bk,\sigma\rangle$ are the Bloch
spinors at zero SO coupling. Diagonalization of the full
single-electron Hamiltonian, $H=H_0+H_{so}$, gives two bands
\begin{equation}
\label{Rashba bands}
    \epsilon_1(\bk)=\epsilon_0(\bk)+\alpha|\bk_\perp|,\quad
    \epsilon_2(\bk)=\epsilon_0(\bk)-\alpha|\bk_\perp|
\end{equation}
($|\bk_\perp|=\sqrt{k_x^2+k_y^2}$), which satisfy the condition
$\epsilon_{1,2}(\bk)=\epsilon_{1,2}(-\bk)$ and additional
symmetries from the point group. Also, the bands touch along the
line $\bk\parallel\hat z$. It is easy to see that one cannot use
pseudospin to label these bands, because the time reversal $K$
transforms the bands into themselves:
$|\bk,n\rangle\to|-\bk,n\rangle$ ($n=1,2$), while the pseudospin
states would be transformed into one another.

To conclude this section, it should be mentioned that the
superconductivity in CePt$_3$Si seems to occur in the presence of
anti-ferromagnetic (AFM) order \cite{exp-CePtSi}, although no data
have been reported on the structure of the magnetic phase or the
magnitude of the staggered moment. Although the symmetry analysis
above was done assuming a paramagnetic normal state, our results
can be easily generalized for an AFM case. For a staggered
magnetization directed along the $z$ axis, the only change one has
to make in the symmetry group is to replace $K$ with
$KT_{\bm{a}}$, which combines the time reversal operation with a
lattice translation. Then, Eq. (\ref{trans K}) is replaced by
\begin{equation}
\label{trans KT}
    KT_{\bm{a}}|\Psi_{\bk}\rangle=
    \left(\begin{array}{cc}
    0 & e^{i\bk\bm{a}} \\
    -e^{-i\bk\bm{a}} & 0 \\
    \end{array}\right)|\Psi_{\bk}\rangle,
\end{equation}
so that $(KT_{\bm{a}})^2|\bk\rangle=-e^{-2i\bk\bm{a}}|\bk\rangle$.

From the expressions (\ref{trans G})--(\ref{trans KT}), one
obtains the transformation rules for the creation operators of
electrons in the Bloch states $|\bk\rangle$. Although there is
some freedom in choosing the phase factors, we will see in Sec.
\ref{sec: OP} that the physically relevant properties are
insensitive to the choice of $\varphi_{\bk}(G)$.

\section{Electronic structure}
\label{sec: Bands}

CePt$_3$Si crystallizes in the same tetragonal structure as
CePt$_3$B, with space group P4mm (No. 99). The lattice parameters
are $a = 4.072$ {\AA} and $c = 5.442$ {\AA}. Ce is at the 1(b)
site ($1/2$, $1/2$, $0.1468$), Pt at the 2(c) site ($1/2$, $0$,
$0.6504$) and at the 1(a) site ($0$, $0$, $0$). The $z$ coordinate
of the latter site was chosen to be zero. Si is at the 1(a) site
($0$, $0$, $0.4118$). These structural parameters all derive from
single-crystal x-ray data \cite{exp-CePtSi}, and can be assumed to
be sufficiently accurate for our purposes.

We calculated the electronic structure of CePt$_3$Si with the
full-potential (FP) linear augmented-plane wave (LAPW) method
\cite{Bla90,Mad01,Sch02}, which is based on density functional
theory \cite{Koh65}. For the exchange and correlation potential we
used the local density approximation \cite{pw92,footnote_gga}
(LDA). We performed a non-magnetic calculation, neglecting the AFM
order observed experimentally below $2.2$ K \cite{exp-CePtSi}. The
muffin-tin radii of the atoms were chosen as $2.11$ $a_0$. A
typical number of plane waves in our basis set was 580. The
electronic ground state was calculated self-consistently on a grid
of 4212 $\mathbf{k}$ points in the entire Brillouin zone.

For the electronic structure of alloys containing heavy elements,
such as Ce and Pt, the SO coupling can in general not be
neglected. In particular, as shown in Sec. \ref{sec: OP}, for the
analysis of the symmetry properties of the superconducting order
parameter it is important to obtain an estimate of the SO
splitting of the bands near the Fermi energy. The SO coupling has
been included in our calculations using the ``second variational
treatment'', as discussed by MacDonald {\it et al} \cite{SO}. In
this approach, first the eigenstates are calculated in the absence
of the SO interaction. Then, the SO interaction is included in a
perturbative way, where the eigenstates up to a certain cut-off
energy, calculated without the SO interaction, are used as the
basis states. In our calculations this cut-off energy was
approximately 22 eV above the Fermi energy.

\begin{figure}
  \includegraphics[angle=270,width=8cm]{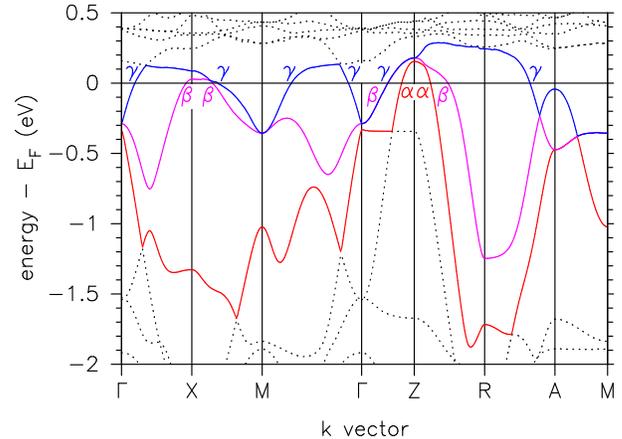}
  \caption{\label{fig_bands1}(Color online) Band structure
  calculated without SO coupling.
  Three bands (labeled $\alpha$, $\beta$, and $\gamma$) cross the Fermi energy.}
\end{figure}

Although the CePt$_3$Si structure doesn't have inversion symmetry,
the band energies  still  satisfy  the relation
$\epsilon_n(\mathbf{k}) = \epsilon_n(-\mathbf{k})$, due to the
time reversal symmetry of the single-electron Hamiltonian (see
Sec. \ref{sec: Symmetry of bands}). Figure \ref{fig_bands1} shows
the band structure of CePt$_3$Si without SO coupling calculated
along some high symmetry lines. The bands were plotted according
to increasing energy. A complete analysis of band crossings based
on the character of eigenfunctions was not carried out. However,
the bands labeled $\beta$ and $\gamma$ do cross between the
symmetry points $X$ and $M$, and so do the $\beta$ and $\gamma$
bands between $R$ and $A$. The $\gamma$ band and the first dotted
band above the Fermi energy have the same symmetry between $M$ and
$\Gamma$, and hence are unlikely to cross there. The labels of the
bands were chosen according to the band index, not the band
character, simply to relate various parts of the Fermi surface to
the bands crossing the Fermi level. In the electronic density of
states (not shown) there is a peak at 0.4 eV above the Fermi
energy, which is due to the unfilled Ce-4f electrons. We found
that the electrons at the Fermi energy are predominantly of Ce-4f
character.

\begin{figure}
  \includegraphics[angle=270,width=8cm]{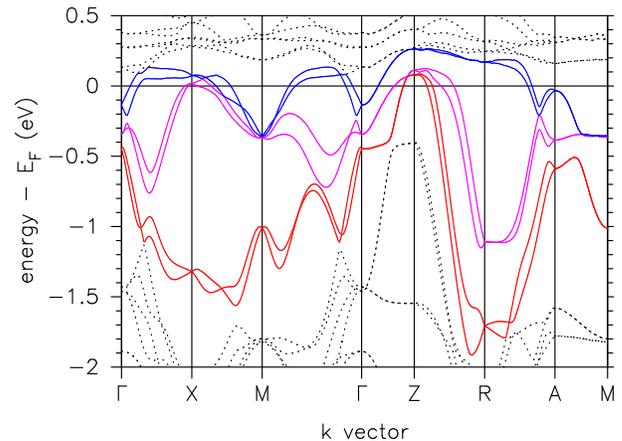}
  \caption{\label{fig_bands}(Color online) Band structure of CePt$_3$Si with
  SO coupling.
  Bands are split due to the SO interaction (cf. Fig. \ref{fig_bands1}).
  The bands that cross the Fermi energy, are plotted as solid lines.}
\end{figure}

Figure \ref{fig_bands} shows the band structure of CePt$_3$Si with
SO coupling \cite{SO}. As in Fig. \ref{fig_bands1} we connected
bands with the same band index, ignoring band crossings. In Fig.
\ref{fig_bands}, the bands near the Fermi energy $\epsilon_F$ are
split by an amount of at most $50-200$ meV. This splitting
vanishes along the lines $\Gamma - Z$ and $M - A$.

\begin{figure}
  \includegraphics[angle=270,width=6cm]{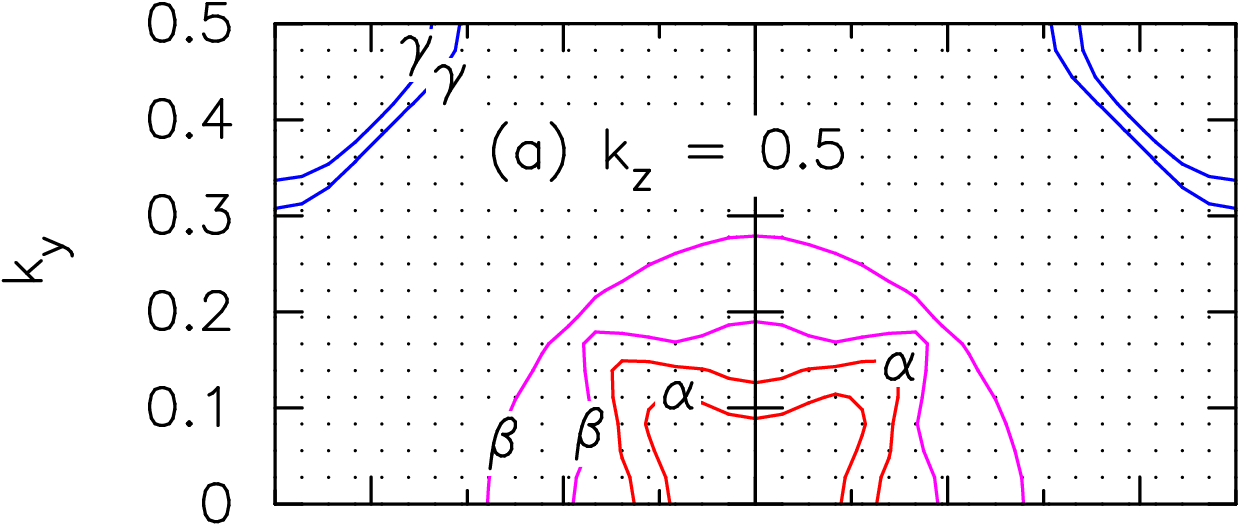}
  \includegraphics[angle=270,width=6cm]{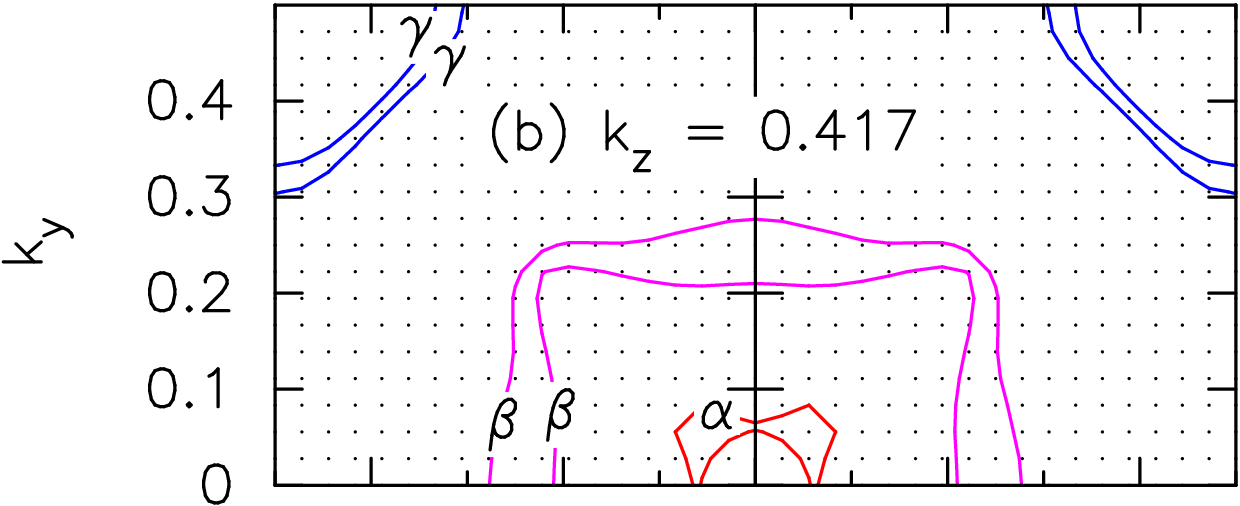}
  \includegraphics[angle=270,width=6cm]{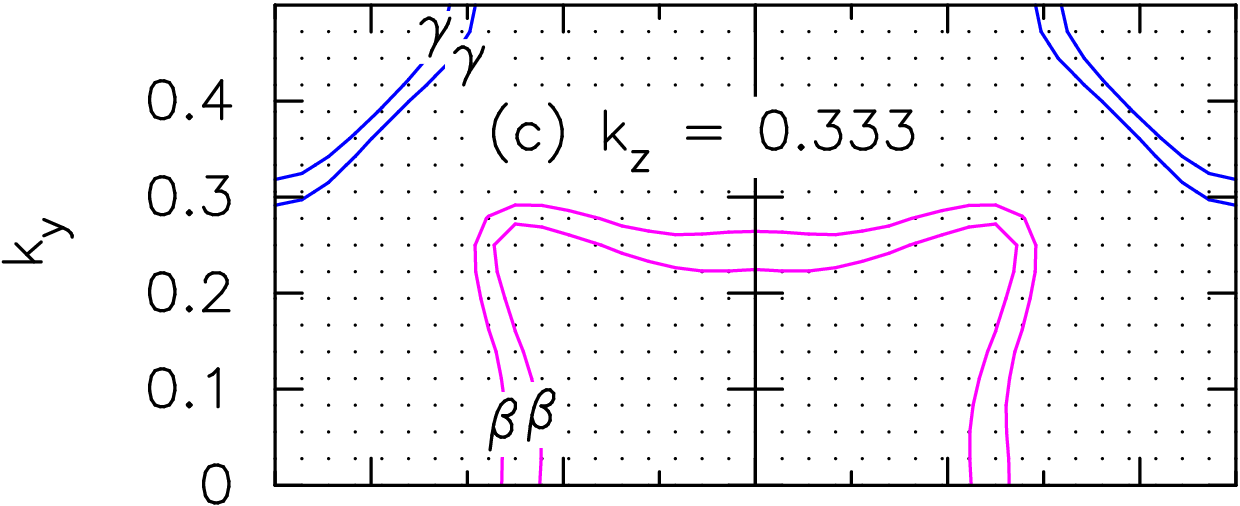}
  \includegraphics[angle=270,width=6cm]{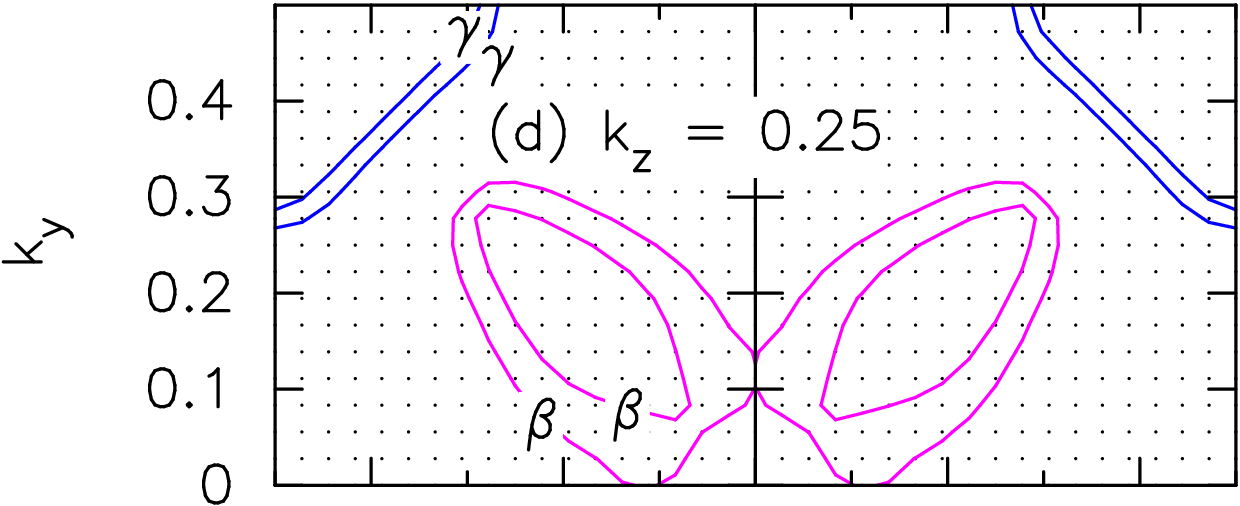}
  \includegraphics[angle=270,width=6cm]{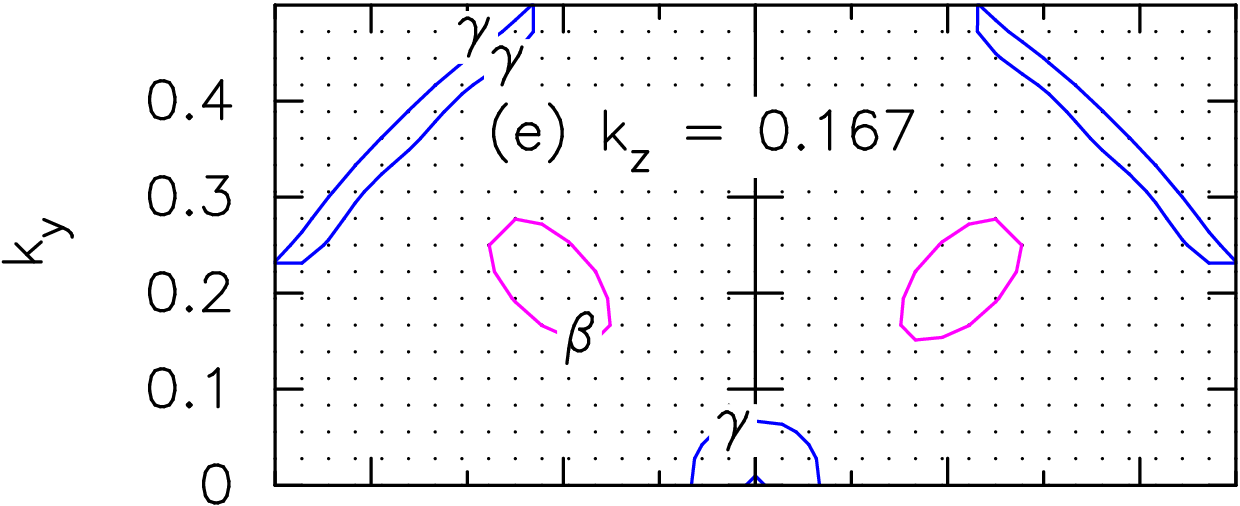}
  \includegraphics[angle=270,width=6cm]{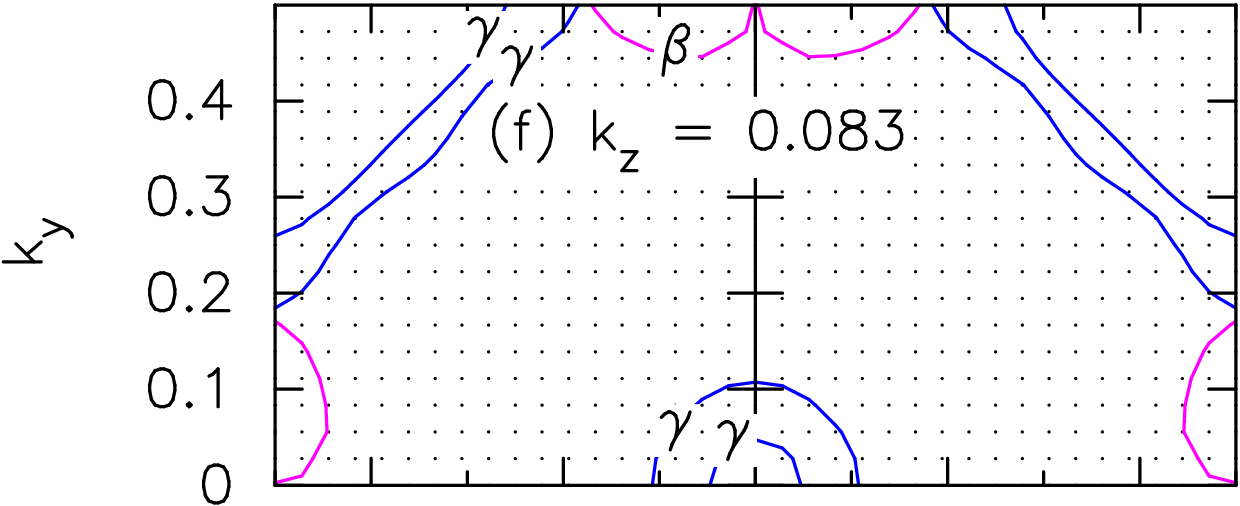}
  \includegraphics[angle=270,width=6cm]{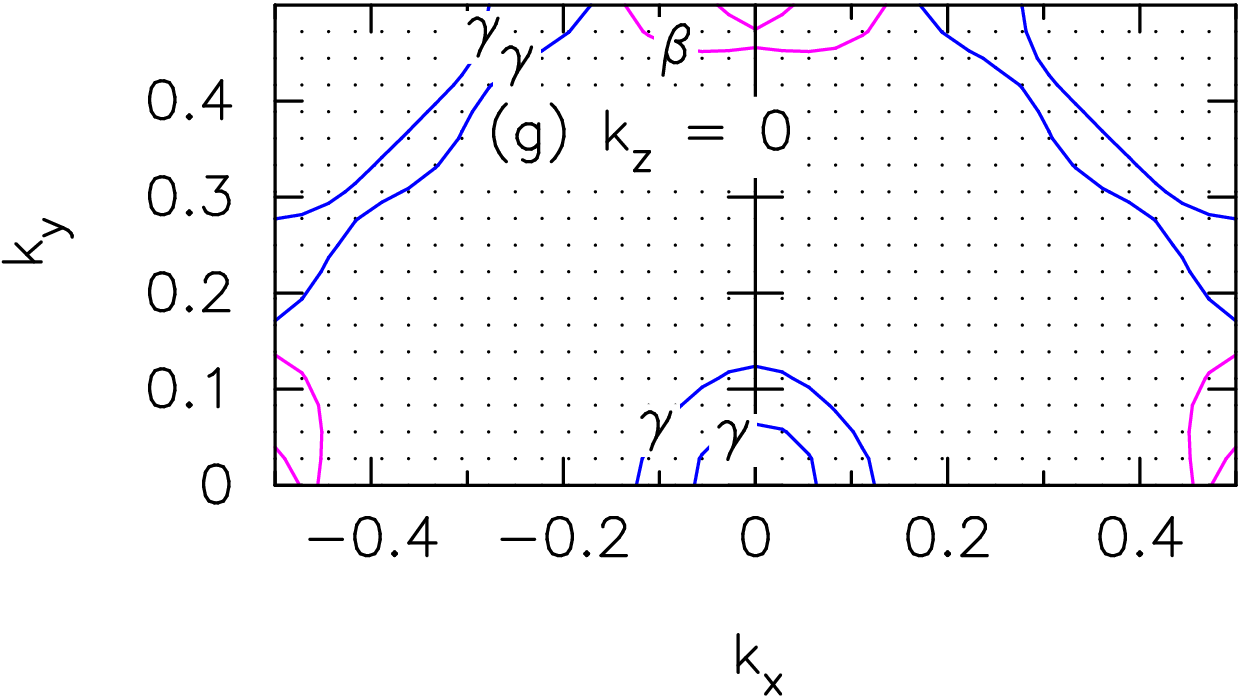}
  \caption{\label{fig_fs}(Color online) Fermi surface of CePt$_3$Si.
  The sub-figures (a)--(g) show cross sections for different values of $k_z$
  ($k_z$ is measured in units of $2\pi/c$, and $k_{x,y}$ are measured in units of
  $2\pi/a$).
  Only one quarter of the Brillouin zone is shown.
  By applying the reflection symmetries in the $k_y = 0$ and $k_z = 0$ planes,
  the Fermi surface in the entire Brillouin zone can be recovered.
  The labels of the sheets correspond to the labels in Fig. \ref{fig_bands1}.
  Dots indicate the $\mathbf{k}$ points for which we calculated the
  band energies.
  The Fermi surface was obtained by interpolating between these points.}
\end{figure}

Figure \ref{fig_fs} shows the cross sections of the Fermi surface
of CePt$_3$Si in the presence of SO coupling. The splitting of the
bands at the Fermi energy due to the SO coupling is prominent in
this figure. The sheets of the Fermi surface labeled $\alpha$ and
$\beta$ are hole-like. The $\gamma$ sheets are electron-like. The
$\alpha$ sheet consists of a feature around the $Z$ point. The
$\beta$ band gives rise to a larger feature around the $Z$ point
and, in addition, to a feature around the $X$ point [and the
symmetry related point $Y = (0,0.5,0)$]. The $\gamma$ bands give
rise to a feature around the line $M - A$, which is almost
dispersionless in $k_z$, as well as to a small feature around the
$\Gamma$ point. We further noted that the $\alpha$ bands
contribute 1.9\% and 3.5\% respectively to the density of states
at the Fermi energy. Similar contributions coming from the SO
split $\beta$ and $\gamma$ bands are 25\% and 45\% and 15\% and
9.0\%, respectively.

\section{Superconducting order parameter}
\label{sec: OP}

\subsection{Symmetry analysis}

The single-electron states $|\bk,n\rangle$ can be used as a basis
for constructing the Hamiltonian which takes into account the
Cooper pairing between electrons. We have $H=H_0+H_{sc}$, where
\begin{equation}
\label{H_0}
     H_0=\sum\limits_n\sum\limits_{\bk}
     \epsilon_n(\bk)c^\dagger_{\bk,n}c_{\bk,n},
\end{equation}
is the free-electron part, with the chemical potential absorbed
into the band energies. As follows from the results of Secs.
\ref{sec: Symmetry of bands} and \ref{sec: Bands}, the electronic
bands $\epsilon_n(\bk)$ are non-degenerate, except along the line
$\bk\parallel\hat z$, and invariant under all operations from the
point group $\mathbf{C}_{4v}$ and also under inversion:
$\epsilon_n(\bk)=\epsilon_n(-\bk)$. The Fermi surface of
CePt$_3$Si consists of several sheets (see Sec. \ref{sec: Bands}),
all of which can in principle participate in the formation of the
superconducting order.

Assuming a BCS-type mechanism of pairing, the interaction between
the band electrons in the Cooper channel can be written in the
following form:
\begin{equation}
\label{H BCS}
    H_{sc}=H^{(1)}_{sc}+H^{(2)}_{sc}+H^{(3)}_{sc},
\end{equation}
where
\begin{eqnarray}
\label{H1}
    H^{(1)}_{sc}=\frac{1}{2}\sum\limits_{n}\sum\limits_{\bk,\bk'}
    V^{(1)}_n(\bk,\bk')c^\dagger_{\bk,n}c^\dagger_{-\bk,n}
    c_{-\bk',n}c_{\bk',n}
\end{eqnarray}
\begin{eqnarray}
\label{H2}
    H^{(2)}_{sc}=\frac{1}{2}\sum\limits_{n\neq m}\sum\limits_{\bk,\bk'}
    V^{(2)}_{nm}(\bk,\bk')c^\dagger_{\bk,n}c^\dagger_{-\bk,n}
    c_{-\bk',m}c_{\bk',m}
\end{eqnarray}
\begin{eqnarray}
\label{H3}
    H^{(3)}_{sc}=\frac{1}{2}\sum\limits_{n\neq m}\sum\limits_{\bk,\bk'}
    V^{(3)}_{nm}(\bk,\bk')c^\dagger_{\bk,n}c^\dagger_{-\bk,m}
    c_{-\bk',m}c_{\bk',n}.
\end{eqnarray}
The potentials $V$ are non-zero only inside the energy shells of
width $\omega_c$ (the cutoff energy) near the Fermi surface.

Treating the Cooper interaction between the electrons with
opposite momenta in the mean-field approximation, we obtain
\begin{equation}
\label{H_sc}
     H_{sc}=\frac{1}{2}\sum\limits_{\bk}\sum\limits_{nm}
     \Bigl[\Delta_{nm}(\bk)c^\dagger_{\bk,n}
     c^\dagger_{-\bk,m}+\mathrm{h.c.}\Bigr].
\end{equation}
Here the diagonal matrix elements $\Delta_{nn}(\bk)$ represent the
Cooper pairs composed of quasiparticles from the same sheet of the
Fermi surface, and the off-diagonal matrix elements
$\Delta_{nm}(\bk)$ with $n\neq m$ represent the pairs of
quasiparticles from different sheets. From the anti-commutation of
the fermionic operators, it follows that $\Delta_{nn}(\bk)$ are
odd functions of $\bk$, but $\Delta_{nm}(\bk)=-\Delta_{mn}(-\bk)$
with $n\neq m$ do not have a definite parity.

A considerable simplification occurs if to assume that the
superconducting gaps are much smaller than the interband energies.
The band structure calculations of Sec. \ref{sec: Bands} show that
typically the SO band splitting $E_{so}$ is of the order of
$50-200\;\mathrm{meV}$ (between the bands derived from the
degenerate spin-up and spin-down bands at zero SO coupling), which
exceeds the superconducting gap by orders of magnitude. In this
situation, the formation of interband pairs is energetically
unfavorable, and the amplitudes $\Delta_{nm}$ with $n\neq m$
vanish. The origin of the suppression of these types of pairing is
similar to the well-known paramagnetic limit of singlet
superconductivity \cite{CC62}: the interband splitting $E_{so}$
cuts off the logarithmic singularity in the Cooper channel, thus
reducing the critical temperature. Although the condition
$E_{so}\gg T_c$ is violated very close to the poles of the Fermi
surface where the spin-split bands touch, the off-diagonal Cooper
pairing in the vicinity of these points is still suppressed due to
the phase space limitations. We also neglect the possibility of
the Cooper pairs having a non-zero momentum, i.e. $\langle
c^\dagger_{\bk+\bq,n}c^\dagger_{-\bk,m}\rangle\neq 0$
[Larkin-Ovchinnikov-Fulde-Ferrell (LOFF) phase] \cite{LOFF}.
Although the critical temperature of the resulting non-uniform
superconducting state can be higher than that of the uniform
state, this would not be sufficient to overcome a large depairing
effect of the SO band splitting.

Thus, the interband pairing (\ref{H3}) can be neglected, and
$\Delta_{nm}(\bk)=\Delta_n(\bk)\delta_{nm}$ \cite{And84}. This is
reminiscent of the situation in ferromagnetic superconductors,
where only the same-spin components of $\Delta$ survive the large
exchange band splitting \cite{FMSC}. It follows from Eqs.
(\ref{trans G})--(\ref{trans K}) and anti-linearity of $K$ that
\begin{eqnarray*}
    &&G(c^\dagger_{\bk,n}c^\dagger_{-\bk,n})G^{-1}=
    c^\dagger_{G\bk,n}c^\dagger_{-G\bk,n}\\
    &&\bar E(c^\dagger_{\bk,n}c^\dagger_{-\bk,n})\bar
    E^{-1}=c^\dagger_{\bk,n}c^\dagger_{-\bk,n}\\
    &&K(\lambda c^\dagger_{\bk,n}c^\dagger_{-\bk,n})K^{-1}=
    -\lambda^*c^\dagger_{-\bk,n}c^\dagger_{\bk,n}\\
    &&=\lambda^*c^\dagger_{\bk,n}c^\dagger_{-\bk,n},
\end{eqnarray*}
where $\lambda$ is an arbitrary constant. We have the following
transformation rules for the order parameter under the elements of
${\cal G}$:
\begin{equation}
\label{Delta transforms} \left.\begin{array}{ll}
     G:& \Delta_n(\bk) \to \Delta_n(G^{-1}\bk),\\
     K:& \Delta_n(\bk) \to \Delta^*_n(\bk).
\end{array}\right.
\end{equation}
Thus, the order parameter components transform like scalar
functions. There is no need for double groups, since $\bar E$ is
equivalent to $E$ when acting on $\Delta$. In the case of an AFM
normal state, $K$ should be replaced with $KT_{\bm{a}}$.

The superconducting order parameter on the $n$th sheet of the
Fermi surface transforms according to one of the irreducible
representations $\Gamma$ of the normal state point group
$\mathbf{C}_{4v}$. It can be represented in the form
\begin{equation}
\label{Delta expansion}
    \Delta_{n,\Gamma}(\bk)=\sum\limits_{i=1}^{d_{\Gamma}}
    \eta_{n,i}\phi_{\Gamma,n,i}(\bk),
\end{equation}
where $d_{\Gamma}$ is the dimension of $\Gamma$,
$\phi_{\Gamma,n,i}(\bk)$ are the basis functions (which are
different on different sheets of the Fermi surface in general),
and $\eta_{n,i}$ are the order parameter components that enter,
e.g., the GL free energy and can depend on coordinates
\cite{Book}. Despite the absence of inversion center in the
crystal, the order parameters $\Delta_n$ have a definite parity,
namely they are all odd with respect to $\bk\to-\bk$. The odd
irreducible representations of $\mathbf{C}_{4v}$ are listed in
Table \ref{table1}. Since $d_\Gamma=1$ or $2$, the order parameter
in each band can have one or two components.

If we neglect the interband pairing described by $H^{(2)}_{sc}$,
see Eq. (\ref{H2}), then the order parameters $\Delta_n$ are
completely decoupled, in particular, they all have different
critical temperatures $T_{c,n}$. However, there is no reason to
expect these interband terms to be small. If they are taken into
account, then all $\Delta_n(\bk)$ are non-zero, so the
superconductivity will be induced simultaneously on all sheets of
the Fermi surface. The simplest way to see how this works is to
use the GL free energy functional, which contains all possible
uniform and gradient terms invariant with respect to ${\cal G}$.
For a one-dimensional representation $\Gamma$ (the generalization
for two-dimensional representations is straightforward), we obtain
the following expression for the uniform terms in the free energy
density:
\begin{equation}
\label{FGLuniform}
   F_{uniform} = \sum\limits_{n,m}A_{nm}(T)\eta_n^*\eta_m+F_4,
\end{equation}
where $F_4$ stands for the terms of fourth order (and higher), and
$A_{ij}$ is a real symmetric matrix. The off-diagonal elements of
$A$ correspond to the interband pairing. The critical temperature
$T_c$ is defined as the maximum temperature at which $A$ ceases to
be positive definite. Below $T_c$ all $\eta_n$ are non-zero and
proportional to a single complex number $\eta$, such that
$\eta_n=\varepsilon_n\eta$, where $\varepsilon_n$ are constants
that can be found by minimizing $F_{uniform}$. Therefore, the
components of a one-dimensional order parameter corresponding to
the representation $\Gamma$ are given by
\begin{equation}
\label{1D}
    \Delta_n(\bk)=\eta\varepsilon_n\phi_{\Gamma,n}(\bk).
\end{equation}
Although the basis functions $\phi$ are different in general, they
all have the same symmetry.

Our phenomenological theory cannot determine which pairing channel
corresponds to the highest critical temperature. In a recent work,
Frigeri {\em et al} \cite {FAKS03} proposed a microscopic model
for superconductivity in a non-centrosymmetric crystal, treating
the SO coupling as a perturbation with a Rashba-type Hamiltonian.
Assuming a strong interband pairing interaction which induces
order parameters of the same amplitude on both sheets of the Fermi
surface [see Eq. (\ref{Rashba bands})], they predicted a certain
gap symmetry for CePt$_3$Si, which seems to correspond to the
two-dimensional $E$ representation in our classification.

\begin{table}
\caption{\label{table1} The character table and the examples of
the odd basis functions for the irreducible representations of
$\mathbf{C}_{4v}$.}
\begin{ruledtabular}
\begin{tabular}{|c|c|c|c|r|}
   $\Gamma$   & $E$ & $C_{4z}$ & $\sigma_x$ &
                    $\phi_{\Gamma}(\bk)$\\ \hline
   $A_1$      & 1   & 1 & 1 &  $k_z$ \\ \hline
   $A_2$      & 1   & 1 & $-1$ &  $(k_x^2-k_y^2)k_xk_yk_z$ \\ \hline
   $B_1$      & 1   & $-1$ & 1 &  $(k_x^2-k_y^2)k_z$ \\ \hline
   $B_2$      & 1   & $-1$ & $-1$ &  $k_xk_yk_z$ \\ \hline
   $E$        & 2   & 0 & 0 & $k_x$\ ,\ $k_y$
\end{tabular}
\end{ruledtabular}
\end{table}

\subsection{Gap zeros}

The symmetry considerations can help find the zeros in the energy
spectrum of Bogoliubov quasiparticles, which are responsible for
peculiarities in the low-temperature behavior of unconventional
superconductors \cite{Book}. The gap structure of the order
parameter belonging to the two-dimensional representation $E$ of
$\mathbf{C}_{4v}$ depends on the superconducting phase, i.e. on
the values of the components $\eta_1$ and $\eta_2$, which in turn
are determined by minimizing the free energy of the
superconductor. In contrast, the gap nodes for the one-dimensional
order parameters are required by symmetry. Although the momentum
dependence of the order parameter is different on different sheets
of the Fermi surface, see Eq. (\ref{1D}), the locations of
symmetry-imposed gap zeros are the same.

One can easily show that the order parameter corresponding to
$A_1$ vanishes on the plane $k_z=0$, so that the energy gap has
lines of nodes at the equators of all sheets of the Fermi surface.
Indeed,
\begin{eqnarray}
\label{C2z}
    C_{2z}\phi_{A_1}(\bk)=\phi_{A_1}(-k_x,-k_y,k_z)\nonumber\\
    =\phi_{A_1}(\bk)=-\phi_{A_1}(-\bk),
\end{eqnarray}
therefore, $\phi_{A_1}(k_x,k_y,0)=0$. In a similar fashion, one
can prove the existence of lines of zeros at $k_z=0$ for all other
one-dimensional representations. Also, the basis functions of
$A_2$ and $B_2$ have lines of zeros at $k_x=0$ or $k_y=0$, while
the basis functions of $A_2$ and $B_1$ have lines of zeros at
$k_x=\pm k_y$. The examples of the basis functions that have only
zeros imposed by symmetry are given in Table \ref{table1}.

From the results of Sec. \ref{sec: Bands} it follows that some of
the sheets of the Fermi surface cross the boundaries of the
Brillouin zone. As seen from Eq. (\ref{C2z}), the order parameters
corresponding to all one-dimensional representations vanish at
$k_z=\pm\pi/c$, i.e. at the top and bottom surfaces of the
Brillouin zone, because $(k_x,k_y,\pi/c)$ and $(k_x,k_y,-\pi/c)$
are equivalent points. In addition, for the same reason the basis
functions of $A_2$ and $B_2$ have lines of zeros at $k_x=\pm\pi/a$
or $k_y=\pm\pi/a$.

The gap nodes for one-dimensional order parameters are present at
the same locations on all sheets of the Fermi surface and can
disappear only if the interband pairing interactions
$H^{(3)}_{sc}$, see Eq. (\ref{H3}), are taken into account.

\subsection{Helical superconducting states}

In addition to the uniform terms (\ref{FGLuniform}), the GL
functional for the order parameter corresponding to a
one-dimensional representation of $\mathbf{C}_{4v}$ also contains
gradient terms
\begin{eqnarray}
\label{FGLgrad1}
   F_{grad,1} &=&
   \sum\limits_{n,m}\Bigl[K^\perp_{nm}(\bm{D}_\perp\eta_n)^*
   (\bm{D}_\perp\eta_m)\nonumber\\
  && +K^z_{nm}(D_z\eta_n)^*(D_z\eta_m)\Bigr],
\end{eqnarray}
where $\bm{D}=\bm{\nabla}+i(2\pi/\Phi_0)\bm{A}$, $\Phi_0=\pi\hbar
c/e$ is the flux quantum, $\bm{A}$ is the vector potential, and
$\bm{D}_\perp=(D_x,D_y)$. The coefficients $K^\perp_{nm}$ and
$K^z_{nm}$ are real symmetric matrices. The terms
(\ref{FGLgrad1}), which are of the second order in $\bm{D}$, are
present in any multi-band tetragonal superconductor with the order
parameter corresponding to a one-dimensional representation of the
symmetry group. However, in the absence of an inversion center,
one can have additional terms in the GL functional, which satisfy
all the necessary symmetry requirements but are of the first order
in gradients \cite{MS94}. In our case, they can be written in the
form
\begin{equation}
\label{FGLgrad2}
    F_{grad,2}=\sum\limits_{n,m}L_{nm}(\eta_n^*D_z\eta_m-
    \eta_m^*D_z\eta_n),
\end{equation}
where $L_{nm}$ is a real anti-symmetric matrix, which is non-zero
only if the interband pairing (\ref{H2}) is present. The terms
(\ref{FGLgrad2}) lead to the possibility that the superconducting
state which appears immediately below $T_c$ can be non-uniform,
even without external magnetic field.

Consider for simplicity only two bands participating in
superconductivity, i.e. $n=1,2$. In this case, the matrix $L$ in
Eq. (\ref{FGLgrad2}) has only one non-zero element:
$L_{12}=-L_{21}=\lambda/2$. The critical temperature for the
superconducting state
\begin{equation}
\label{helical}
    \eta_1(\br)=\eta_{1,0}e^{iqz},\quad \eta_2(\br)=\eta_{2,0}e^{iqz}
\end{equation}
is determined by the stability condition of the quadratic terms
(both uniform and gradient) in the GL functional towards formation
of a state with non-zero $\eta_{n,0}$. From Eqs.
(\ref{FGLuniform},\ref{FGLgrad1},\ref{FGLgrad2}), one has the
following equation for $T_c(q)$:
\begin{equation}
\label{Tcq eq}
    \det\left|\begin{array}{cc}
        A_{11}+K_{11}q^2 & A_{12}+K_{12}q^2+i\lambda q\\
        A_{12}+K_{12}q^2-i\lambda q & A_{22}+K_{22}q^2
    \end{array}\right|=0,
\end{equation}
where $A_{11}(T)=a_1(T-T_1)$ and $A_{22}(T)=a_2(T-T_2)$, with
$T_1$ and $T_2$ having the meaning of the critical temperatures
for the bands 1 and 2 respectively in the absence of any interband
coupling. The phase transition temperature is obtained by the
maximization of $T_c(q)$ with respect to $q$. It is easy to show
that the maximum critical temperature corresponds to a state with
$q\neq 0$ (a ``helical'' state \cite{MS94}) if the following
condition is satisfied:
\begin{equation}
\label{condition q}
    \lambda^2+2A_{12}K_{12}-A_{11}(T_{c,0})K_{22}-A_{22}(T_{c,0})K_{11}>0,
\end{equation}
where $T_{c,0}=T_c(q=0)$. However, even if this condition is
violated and the phase transition occurs from the normal state to
a uniform superconducting state, there remains a possibility that
this uniform state becomes unstable towards the formation of a
helical state at a lower temperature. To find this instability,
one has to include non-linear terms in the free energy. Because of
a large number of the phenomenological parameters in the GL
functional with the higher-order terms, the phase diagram of this
system is quite rich \cite{MS94}. In particular, there exist
various types of helical phases with $q\neq 0$, separated from one
another and from the uniform phase by additional phase transitions
below $T_c$.

It should be emphasized that the origin of the helical
superconducting states is different from that of the LOFF
non-uniform states \cite{LOFF}. In terms of the GL functional, the
LOFF state corresponds to the sign change of the second-order
gradient term at some values of the parameters (e.g. of the
external magnetic field), while our helical instability occurs
because of the presence of the first-order gradient terms.

\section{Conclusion}
\label{sec: Concl}

We have shown that the order parameter in a non-centrosymmetric
superconductor with strong spin-orbit coupling has only intra-band
components and is always odd with respect to $\bk\to-\bk$, which
is a consequence of Pauli principle. This should be contrasted to
the case of zero spin-orbit coupling, in which the bands are
two-fold degenerate. In that limit, one cannot separate the odd
and even components of $\Delta$ because of the lack of inversion
symmetry, so the order parameter does not have a definite parity.

The Fermi surface of CePt$_3$Si consists of three pairs of sheets,
$\alpha$, $\beta$, and $\gamma$, each split by the spin-orbit
coupling. Our band structure calculations reveal that the states
at the Fermi level are predominantly of Ce-4\textit{f} character.
These states are affected strongly by spin-orbit coupling, which
leads to the band splitting energy as high as $50-200$ meV. The
splitting vanishes along the $\Gamma$ -- \textit{Z} and \textit{A}
-- \textit{M} symmetry lines. By far the biggest contribution to
the density of states at the Fermi level comes from the $\beta$
sheets.

Although the large value of the spin-orbit band splitting excludes
the superconducting states that correspond to the pairing of
electrons from different sheets of the Fermi surface, one can
expect that the interband pair scattering will induce gaps of the
same symmetry on all sheets of the Fermi surface. Possible gap
structure of CePt$_3$Si depends on the dimensionality of the order
parameter. If the order parameter corresponds to a one-dimensional
representation of the group $\mathbf{C}_{4v}$, then the gap has
line nodes where the Fermi surface crosses the high-symmetry
planes or the boundaries of the Brillouin zone: at
$k_z=0,\pm\pi/c$ for all 1D representations; at $k_x=0,\pm\pi/a$
and $k_y=0,\pm\pi/a$ for $A_2$ and $B_2$; at $k_x=\pm k_y$ for
$A_2$ and $B_1$.

The presence of the gap nodes would manifest itself, e.g. in a
power-law behavior of thermodynamic and transport characteristics
at $T\to 0$. Although the gap symmetries on all sheets of the
Fermi surface should be the same, their magnitudes may be
different. The experimental data, e.g. a reduced value of the
specific heat jump at $T_c$ \cite{exp-CePtSi}, indicate that only
some parts of the Fermi surface have non-zero superconducting
gaps, while others remain normal. If this is indeed the case, then
the specific heat would drop as $C(T)/T\propto \mathrm{const}\,+a
T$ at low temperatures (with the constant contribution coming from
the normal sheets of the Fermi surface), which seems to agree with
the experimental data of Ref. \cite{exp-CePtSi} in zero field.
More detailed information about the pairing symmetry can be
obtained only if the precise location of the line nodes is known.

The absence of inversion symmetry can also have interesting
consequences for the spatial structure of the superconducting
phase. We showed that, in contrast to the centrosymmetric case,
the Ginzburg-Landau free energy can now contain additional terms
which are of the first order in gradients. Such terms can make the
superconducting phase unstable towards the formation of a
non-uniform (helical) state even at zero magnetic field. For the
order parameters corresponding to the one-dimensional
representations of $\mathbf{C}_{4v}$, this possibility exists only
if the interband pairing is taken into account.

\section*{Acknowledgements}

The authors thank B. Mitrovic for useful discussions, and also D.
Agterberg, E. Bauer, and especially V. Mineev for stimulating
correspondence. This work was supported by the Natural Sciences
and Engineering Research Council of Canada.

\newpage

\end{document}